\documentclass[aps,prev,twocolumn,preprintnumbers,floatfix,nofootinbib,longbibliography]{revtex4-1}

\usepackage[T1]{fontenc} 
\usepackage[utf8]{inputenc}
\usepackage{physics}
\usepackage{subfigure}

\usepackage{amssymb,graphicx}
\usepackage{mathrsfs}
\usepackage{verbatim}
\usepackage{color}
\usepackage[dvipsnames]{xcolor}
\usepackage{multirow}
\usepackage{amsmath}
\usepackage{subfigure}
\usepackage{slashed} 
\usepackage{float}
\usepackage[normalem]{ulem}
\usepackage{sidecap}
\usepackage{hyperref}
\usepackage{cancel}
\usepackage{enumerate}
\usepackage{array}
\hypersetup{pdfstartview=FitV,colorlinks=true,linkcolor=blue,citecolor=red,filecolor=black,urlcolor=blue}

\newcommand{\aend}{a_{\rm end}}
\newcommand{\arh}{a_{\rm RH}}
\newcommand{\rhorh}{\rho_{\rm RH}}
\newcommand{\rhoend}{\rho_{\rm end}}
\newcommand{\phiend}{\phi_{\rm end}}
\newcommand{\trh}{T_{\rm RH}}
\newcommand{\mchi}{m_\chi}
\newcommand{\beq}{\begin{equation}}
\newcommand{\eeq}{\end{equation}}
\newcommand{\bea}{\begin{equation}}
\newcommand{\eea}{\end{equation}}
\newcommand{\be}{\begin{equation}}
\newcommand{\ee}{\end{equation}}
\newcommand{\ba}{\begin{eqnarray}}
\newcommand{\ea}{\end{eqnarray}}

\def\trh{T_{\rm RH}}

\usepackage[parfill]{parskip}
\begin{document}\sloppy

 \preprint{UMN--TH--4401/24, FTPI--MINN--24/21}

\vspace*{1mm}

\title{The Role of the Curvaton Post-Planck}
\author{Gongjun Choi}
\email{choi0988@umn.edu}
\author{Wenqi Ke}
\email{wke@umn.edu}
\author{Keith A. Olive}
\email{olive@umn.edu}
\vspace{0.5cm}
\affiliation{William I. Fine Theoretical Physics Institute, School of
 Physics and Astronomy, University of Minnesota, Minneapolis, MN 55455,
 USA}

\date{\today}

\begin{abstract} 
The expected improvements in the  precision of inflationary physics observables including the scalar spectral index $n_{s}$ and the tensor-to-scalar ratio $r$ will reveal more than just the viability of a particular model of inflation. In the presence of a curvaton field $\chi$, supposedly dead models of inflation can be resurrected as these observables are affected by curvaton perturbations. For currently successful models, improved constraints will enable us to constrain the properties of extra decaying scalar degrees of freedom produced during inflation. In this work, we demonstrate these diverse uses of a curvaton field with the most recent constraints on ($n_{s},r$) and two exemplary inflation models, the Starobinsky model, and a model of new inflation. Our analysis invokes three free parameters: the curvaton mass $m_{\chi}$, its decay rate $\Gamma_{\chi}$ the reheating temperature $T_{\rm RH}$ produced by inflaton decays. We systematically analyze possible post-inflationary era scenarios of a curvaton field. By projecting the most recent CMB data on ($n_{s},r$) into this parameter space, we can either set constraints on 
the curvaton parameters from successful models of inflation (so that the success is not spoiled) or determine the parameters which are able to save a model for which $n_{s}$ is predicted to be below the experimental data. We emphasize that the initial value of $\langle \chi^2 \rangle \propto H^4/m_\chi^2$ produced during inflation is determined from a stochastic approach and thus not a free parameter in our analysis. We also investigate the production of local non-Gaussianity $f_{NL}^{(\rm loc)}$ and apply current CMB constraints to the parameter space.  
Intriguingly, we find that a large value of $f_{NL}^{(\rm loc)}$ of $\mathcal{O}(1)$ can be produced for both of the two representative inflation models.

\end{abstract}

\maketitle

\setcounter{equation}{0}

\section{Introduction}
Every model of inflation is characterized by its prediction of the cosmic microwave background (CMB) anisotropy spectrum and in particular, the spectral index $n_{s}$ of scalar perturbations and the tensor-to-scalar ratio $r$. As a consequence, constraints on $n_{s}$ and $r$ are used to triage models of inflation. For instance, the original Starobinsky model~\cite{Staro} predicts  $r\sim0.004$ for 55 $e$-folds, consistent with experimental upper limit of $r < 0.036$ from a combination of Planck and and BICEP/{\it Keck} \cite{Planck:2018jri,BICEP2021,Tristram:2021tvh}. In contrast, models of chaotic inflation \cite{chaotic} characterized by a quartic potential predict $r = 0.27$ and are no longer considered viable. 
As such, one may be tempted to rule out parameter space of a model that violates the most recent constraints on $r$ and $0.956 < n_s < 0.973$ at 95\% C.L. 

Such exclusions are justified provided that the observables $n_{s}$ and $r$ receive contributions  from a single inflaton perturbation during inflation. This conclusion, however, can be hasty in cases for which $n_{s}$ and $r$ have contributions from the perturbations in other fields. The presence of a scalar spectator field $\chi$ during inflation can affect the predictions of $n_s$ and $r$ if its effective mass is smaller than the Hubble scale during inflation, $m_{\chi}\ll H_{\rm I}$, and it decays after inflaton reheating. The isocurvature perturbations in $\chi$ and a fraction of the total energy contributed by $\chi$ give rise to contributions to $n_{s}$ and $r$. Thus, whenever inflationary models are extended to include the aforementioned light spectator scalar field, or curvaton~\cite{Enqvist:2001zp,Lyth:2001nq,Moroi:2001ct}, there can be a change to viable parameter space of models consistent with constraints from CMB observables. 

When constraints on $n_{s}$ and $r$ cannot be explained by a single field inflation model, the curvaton becomes particularly useful: the excluded parameter space can be resurrected with the aid of the curvaton. But, even for inflation models which can account correctly for $n_{s}$ and $r$, it is still informative to determine the modification to $n_{s}$ and $r$ in the presence of $\chi$ and constrain the viable parameter space of potential candidates for curvaton fields. 

In some sense, introducing the curvaton seems to complicate our understanding for the early universe physics because the curvaton makes the landscape of working scenarios richer. Put another way, for a given set of observables, ($n_{s},r$), it is not possible to constrain the curvaton sector unless we clearly specify the inflaton sector. In this case,  an additional observable is particularly useful. Namely, the prediction of the local-type non-Gaussianity $f_{NL}^{(\rm loc)}$.  In single field inflationary models, $f_{NL}^{(\rm loc)} \sim \mathcal{O}(10^{-2})$~\cite{Acquaviva:2002ud,Maldacena:2002vr}, and the ambiguity may be eliminated when a non-zero value of $f_{NL}^{(\rm loc)}$ is detected as a curvaton can produce a significant amount of non-Gaussianity in the primordial density perturbation~\cite{Bartolo:2004ty,Lyth:2005fi,Boubekeur:2005fj}. The current constraint on $f_{NL}^{(\rm loc)}$ from the Planck 2018 is $f_{NL}^{(\rm loc)}=-0.9\pm5.1$~\cite{Planck:2019kim} at $68\%$ C.L..

In what follows, we first review the CMB observables in single inflationary models and how they are affected by the presence of a curvaton. In section \ref{sec:eom}, we outline the coupled set of equations of motion for the inflaton, curvaton and radiation produced by both inflaton and curvaton decay. Since we assume $\mchi \ll H_{\rm I}$ during inflation, we use the quantum fluctuations in $\chi$ produced during inflation to determine the initial conditions for the curvaton. As a consequence, we are able to describe the possible scenarios with three parameters: the curvaton mass, $m_{\chi}$, and decay rate, $\Gamma_{\chi}$, and the inflaton reheating temperature, $T_{\rm RH}$ (or equivalently the inflaton decay rate $\Gamma_{\phi}$). We assume the the scale of inflation is fixed by the normalization of the CMB anisotropy spectrum.  These scenarios are described in detail in Section \ref{sec:scenarios} and are distinguished by the order of events leading to curvaton oscillations and decays relative to inflaton decays. These results are applied to two very different models of inflation in Section \ref{sec:app}. Our conclusions are found in Section \ref{sec:disc}.

\section{Cosmological Observables}

One of the most important attributes of inflationary theory is the fact that we can identify four observables
which can test the theory's viability. One of the key predictions of inflation is the production of curvature perturbations. 
The normalization of the power spectrum, $P_{\zeta}(k)$, of the curvature perturbation is fixed by
the observations of the CMB anisotropies. Expanding the scalar power spectrum as 
\beq
\ln P_\zeta(k) = \ln A_s + (n_s-1) \ln(k/k_*) + \cdots \,.
\eeq
Planck \cite{Planck:2018jri} has determined $A_{s}\simeq (2.10 \pm 0.03) \times10^{-9}$ 
 at the CMB pivot scale $k_{*}=0.05{\rm Mpc}^{-1}$.
For a given model of inflation, i.e. scalar potential for the inflaton, this measurement amounts to fixing the scale of inflation or the Hubble parameter during inflation, $H_{\rm I}$. 

The next term in the expansion of the power spectrum is the scalar tilt, $n_s$.  From our understanding of large scale structure, it was expected that the power spectrum is approximately flat with $n_s \approx 1$ \cite{Harrison:1969fb,Zeldovich:1972zz} and first calculated for the Starobinsky model \cite{Staro} by  
Mukhanov and Chibisov \cite{Mukhanov:1981xt} who found a slightly red tilt, i.e. $n_s < 1$. 
This too has been measured to high precision by Planck \cite{Planck:2018jri} to be $n_s = 0.9649 \pm 0.0042$ and has played an important role in discerning between different inflation models. 

 In addition to the scalar power spectrum, inflation models predict a spectrum of tensor perturbations.  The ratio of the 
 tensor-to-scalar perturbations, $r$, at the same pivot  $k_{*}$, 
 has not been measured, yet a strong upper limit of $r < 0.036$ has been set \cite{BICEP2021,Tristram:2021tvh}.
 This upper limit has also played an important role in constraining models of inflation as discussed above. In principle, the tensor tilt $n_T$, is also a predicted observable, but to date as the value of $r$ has not been measured, there is clearly no measurement or limit on $n_T$.

Finally there is the possibility (particularly in multi-field models of inflation) that the perturbation spectrum is not perfectly Gaussian. The degree of non-Gaussianity can be parameterized by the coefficient of the CMB bispectrum,  $f_{\rm NL}^{(\rm loc)}$. Planck has not seen evidence of non-Gaussianity and limits $f_{\rm NL}^{(\rm loc)} = -1 \pm 5$ \cite{Planck:2019kim}. 

The inflationary observables, can be easily computed in single field models from the inflaton potential. 
The slow-roll parameters are defined as
\be
\epsilon_\varphi\equiv \frac{1}{2}M_P^2\left(\frac{V_\varphi}{V}\right)^2,\quad \eta_\varphi\equiv M_P^2\left(\frac{V_{\varphi\varphi}}{V}\right)\, ,
\ee
where $V$ is the total scalar potential and the subscripts denote the derivatives with respect to the subscript fields. For single field inflation models, $V$ is just the inflaton potential\footnote{These definitions are general so that they may be applied to multiple fields such as the inflaton, $\phi$ and curvaton $\chi$ with $\varphi\equiv \phi,\chi$.}. From these,  
we can calculate
\beq
A_s \; \simeq \; \frac{V_*}{24 \pi^2 \epsilon_* M_{P}^4 } \, ,
\label{As}
\eeq
\beq
n_s  = n_* \;  \simeq \; 1 - 6 \epsilon_* + 2 \eta_*\ \, ,
\label{ns}
\eeq
and 
\beq
r = r_* \simeq 16 \epsilon_* \, .
\label{r}
\eeq
In each case the subscript $*$ refers
to evaluating the slow-roll parameters for a single field model with the potential (and its derivatives) evaluated at the field value $\phi_*$. 

Inflation ends when the accelerated expansion of the universe stops,  that is when $\ddot{a} = 0$ and $a=\aend$, where $a$ is the cosmological scale factor. Inflaton oscillations (and the matter dominated era) begin at $a= \aend$. 
In terms of the (potential) slow-roll parameters, we have $\epsilon_\phi \simeq \left(1+\sqrt{1-\eta_\phi/2}\right)^2$. The number of $e$-folds between the CMB pivot scale and the end of inflation can be computed by
\begin{equation}
    N_*\simeq-\frac{1}{M_P^2}\int^{\phi_\text{end}}_{\phi_*}\frac{V}{V_\phi}d\phi \, .
\end{equation}
Thus once the inflation model is specified,
for a given number of $e$-folds, the three observables in Eqs.~(\ref{As})-(\ref{r}) are calculable ($f_{\rm NL}^{(\rm loc)}$ is negligible in this case). The relation between $N_*$ and $n_s$ can be improved once the inflation model is extended to include reheating as discussed in more detail below.

The presence of a massive (unstable) scalar field which comes to dominate (or close to dominating) the energy density, 
may affect the above predictions of the inflationary model. Such a field, 
a curvaton, picks up isocurvature fluctuations which (upon decay) alter the CMB observables \cite{Enqvist:2001zp,Lyth:2001nq,Moroi:2001ct,Wands:2002bn,Lyth:2002my,Dimopoulos:2002kt,Moroi:2005kz,Moroi:2005np,Linde:2005yw,Malik:2006pm,Sasaki:2006kq,Enqvist:2009zf,Mazumdar:2010sa,Fonseca:2012cj,Enqvist:2012xn,Byrnes:2014xua,Fujita:2014iaa,He:2015msa,Smith:2015bln,Byrnes:2016xlk,Torrado:2017qtr,Kumar:2019ebj,Lodman:2023yrc}. 
In the remainder of the section, we
assume the presence of a massive scalar field, $\chi$, with potential $V(\chi)$.

The non-adiabatic part of the perturbation in $\chi$ is quantified by \cite{Fonseca:2012cj}
\beq
S_{\chi} \equiv 3(\zeta_{\chi}-\zeta_{\phi}) \, ,
\label{eq:Schi}
\eeq
where $\zeta_\phi$ and $\zeta_\chi$ are curvature perturbations in $\phi$ and $\chi$ respectively. 
Then in the approximation of $\chi_{\rm osc}\simeq\chi_{\rm inf}+\delta\chi_{*}$,
where $\chi_{\rm inf}$ is the initial background field value of $\chi$ at the end of inflation (discussed in more detail in the next section), we can write \cite{Fonseca:2012cj}
\beq
S_{\chi} \simeq  2\frac{\delta\chi_{*}}{\chi_{\rm inf}} \, .
\eeq
Then for $\delta \chi_* = H_{\rm end}/2\pi$ \footnote{The value of the Hubble parameter at the end of inflation, $H_{\rm end}$ is slightly below its value during inflation, $H_{\rm I}$ though we do not keep this distinction here.} and assuming $\mchi \ll H_{\rm end}$,  $\chi_{\inf} = \sqrt{3/8 \pi^2}H_{\rm end}^2/\mchi$ (see below), we have 
\ba
S_{\chi}&\simeq & 2\left(\frac{H_{\rm end}}{2\pi}\right)\sqrt{\frac{8\pi^{2}}{3}}\frac{\mchi}{H_{\rm end}^{2}}\cr\cr
&\simeq&2\sqrt{\frac{2}{3}}\frac{\mchi}{H_{\rm end}}\,,
\label{eq:Schi2}
\ea
From (\ref{eq:Schi2}), we obtain for the curvaton power spectrum
\be
P_{S_{\chi}}\simeq\frac{8}{3}\left(\frac{\mchi}{H_{\rm end}}\right)^{2}\,.
\label{eq:PSchi}
\ee

When the curvaton decays, the isocurvature perturbations are transferred to curvature perturbation in the radiation
with the total perturbation given by~\cite{Fonseca:2012cj}
\beq
\zeta = R_\chi \zeta_\chi + (1-R_\chi) \zeta_\phi\, ,
\label{eq:zetatot}
\eeq
with 
\beq
R_{\chi}=\frac{3\rho_{\chi}}{4 \rho_{\rm R} + 3\rho_{\chi}}\biggr\rvert_{d\chi} \, ,
\eeq
evaluated when $\chi$ decays ($d\chi$). Here, $\rho_{\rm R}$ is the existing radiation density from inflaton decays. It will be useful to compute 
$R_\chi$ below from the ratio of the curvaton to the radiation density, and we define the factor 
$x\equiv (\rho_\chi/\rho_r)|_{d\chi}$ so that 
\be
R_\chi = \frac{3 x}{4+3x} \,,
\label{Rofx}
\ee 
and the cases $x<1$ ($x>1$) 
will represent radiation (matter) domination scenarios at the time of curvaton decay. Then, from (\ref{eq:Schi}) and (\ref{eq:zetatot})
\beq
\zeta = \zeta_\phi + \frac13 R_\chi S_\chi\,,
\eeq
and since the fluctuations in $\phi$ and $\chi$ are uncorrelated, we can write
\beq
P_{\zeta} =P_{\zeta_{\phi}}+\frac{R_{\chi}^{2}}{9}  P_{S_{\chi}}\,.
\eeq
As before, the total spectrum is normalized by $P_\zeta (k_*) = A_s$.
If we define the ratio of the curvaton contribution to the power spectrum relative to the total as
\beq
\omega_{\chi}\equiv\frac{R_{\chi}^{2}P_{S_{\chi}}}{9P_{\zeta}}\,,
\label{eq:defs}
\eeq
we can rewrite the CMB observables as
\beq
n_{s}= n_* + \omega_\chi \left(4 \epsilon_{*}+2\eta_{\chi}- 2\eta_{*}\right) \, .
\label{eq:netns}
\eeq
The contribution from the curvaton to the spectral index is:
\be
\delta n_s=\omega_\chi(4\epsilon_*+2\eta_\chi-2\eta_*)=\frac{16}{27}\frac{R_\chi^2}{A_s}\left(\frac{\mchi}{H_{\rm end}}\right)^{2}(2\epsilon_*+\eta_\chi-\eta_*) \,,
\label{eq:deltans}
\ee
where we used (\ref{eq:PSchi}) and (\ref{eq:defs}) for the second equality.

In addition, $\omega_{\chi}$ in (\ref{eq:defs}) enters in the modified tensor-to-scalar ratio
\beq
r=r_{*}(1-\omega_{\chi})
\, ,
\label{deltar}
\eeq
and the local-type non-Gaussianity parameter is
\beq
f_{NL}^{(\rm loc)}=\left(\frac{5}{4R_{\chi}}-\frac{5}{3}-\frac{5}{6}R_{\chi}\right)\omega_{\chi}^{2}\,.
\label{eq:parameter}
\eeq

\section{Cosmological evolution of the inflaton and curvaton}
\label{sec:eom}

Given an inflationary model, we will 
assume a very simple form for the curvaton potential, 
\beq
V(\chi) = \frac12 m_\chi^2 \chi^2 \, .
\eeq
In all of our analyses below, $m_{\chi}$ will refer to the effective mass of the curvaton which includes all contributions from self interactions or interactions with other fields. For simplicity, we assume however that $\mchi$ is constant.
As the curvaton is required to decay (if stable, $\chi$ may comprise dark matter\footnote{The cosmological evolution of stable spectator fields have been studied in \cite{Turner:1987vd,Peebles:1999fz,Enqvist:2014zqa,Nurmi:2015ema,Bertolami:2016ywc,Alonso-Alvarez:2018tus,Markkanen:2018gcw,Tenkanen:2019aij,Choi:2019mva,Cosme:2020nac,Lebedev:2022cic,cgkmov}.}, but cannot affect the inflationary observables and is subject to constraints from isocurvature fluctuations), we must also couple the curvaton to Standard Model fields, but we are only interested in its lifetime or decay rate, $\Gamma_\chi$.  In addition the cosmological history will depend on the reheating temperature, $\trh$, which is determined by the inflaton couplings to Standard Model fields
and can be parameterized by the inflaton decay rate, $\Gamma_\phi$.
Thus our study can be cast in terms of three parameters which we take to be a priori as free:
$m_{\chi}, \Gamma_\chi$, and $\trh$. 

In principle, to study the evolution of the curvaton, we must also specify its initial field value, $\chi_i$. 
During inflation, all scalar fields are subject to quantum fluctuations \cite{fluc,Racco:2024aac}, which grow linearly in time up to an asymptotic value determined by $H_I$ and the mass of the scalar. For
$\mchi \ll H_I$, the fluctuations characterized by $\langle \chi^2 \rangle$ can be much larger than $H_I^2$ and the long wavelength modes of these fluctuations that obey the classical equations of motion \cite{longw} are indistinguishable from a homogeneous scalar field. 

Typically when $\phi = \phi(\aend) = \phiend$, the energy density in the inflaton given by $\rho_\phi(\aend) = \rhoend = \frac32 V(\phiend)$ is somewhat less than the inflaton energy density during inflation.  Hence the Hubble parameter, $H_{\rm end} \lesssim H_{\rm I}$.  
Therefore, we will assume that the initial value of $\chi$ is given by \cite{Linde:2005yw,Torrado:2017qtr,cgkmov}
\be
\chi_{\rm inf}^{2} = \langle \chi^2 \rangle \simeq \frac{3}{8\pi^{2}}
\left(\frac{H_{\rm end}}{\mchi}\right)^{2}H_{\rm end}^{2}\, .
\label{eq:sigmainf}
\ee
While one can choose to (arbitrarily) take initial field values larger than that given in (\ref{eq:sigmainf}), the quantum fluctuations produced during inflation provide a lower bound to the initial scalar field value. In all that follows, we assume the relation between $\chi_{\rm inf}$ and $\mchi$ in (\ref{eq:sigmainf}) and the value of $H_{\rm end}$ specified by the choice of the model of inflation.

The three parameters ($m_{\chi}, \Gamma_\chi$, and $\trh$)
will determine the order of events when $\chi$ begins to oscillate, denoted as $a_\chi$, when $\chi$ decays at $a_{d\chi}$ and reheating at $\arh$. We assume that all three events take place after the end of inflation, i.e., $a_\chi, a_{d\chi},\arh > \aend$. We will also assume that $a_\chi < a_{d\chi}$.\footnote{This amounts to assuming a perturbative decay of $\chi$. If $\Gamma_\chi$ is parameterized by a Yukawa-like interaction with $\Gamma_\chi = (y_\chi^2/8\pi) m_\chi$, $a_\chi < a_{d\chi}$ is equivalent to assuming $\Gamma_\chi < \mchi$.}

The equations of motion for the two scalars are
\begin{equation}
    \label{eq:eomphi}
    \ddot{\phi} + 3H \dot{\phi} + \Gamma_\phi \dot{\phi} + \frac{dV(\phi)}{d\phi} \; = \; 0 \, ,
\end{equation}
\begin{equation}
    \label{eq:eomS}
    \ddot{\chi} + 3H \dot{\chi} + \Gamma_\chi \dot{\chi} + \mchi^2 \chi \; = \; 0 \, .
\end{equation}
In terms of the energy densities, we can write
\beq
\dot{\rho}_\phi + 3H \rho_\phi + \Gamma_\phi \rho_{\phi} = 0 \, ,
\label{rhophidot}
\eeq
\beq
\dot{\rho}_\chi + 3 H \rho_\chi + \Gamma_\chi \rho_{\chi} = 0 \, ,
\eeq
\beq
\dot{\rho}_{\rm R} + 4 H \rho_{\rm R} = \Gamma_\phi \rho_\phi + \Gamma_\chi \rho_\chi \, ,
\eeq
together with 
\beq
\rho_\phi + \rho_\chi + \rho_{\rm R} = 3 H^2 M_P^2 \, ,
\eeq
where $\rho_{\rm R}$ is the energy density 
of radiation produced by inflaton (or curvaton) decay, $H = {\dot a}/a$, and $M_P = 2.4 \times 10^{18}$~GeV is the reduced Planck mass. We have also assumed that the equation of state for the inflaton and curvaton (post-inflation) during their oscillatory phase is pressureless ($P_{\varphi}/\rho_{\varphi}=0$, $\varphi=\phi,\chi$) which is valid if the inflaton oscillations are governed by a quadratic term about its potential minimum. 

Prior to inflaton reheating, the expansion rate is dominated by the energy density in inflaton oscillations and the solution to (\ref{rhophidot}) for $\Gamma_\phi \ll H$ is
\be
\rho_\phi(a) = \rho_{\rm end} \left( \frac{\aend}{a} \right)^3 \,.
\label{eq:rhophi}
\ee
We also assume that the inflationary sector is coupled to the Standard Model so as to account for reheating.
As the inflaton starts to decay, the radiation density quickly rises to a maximum at  
$a = a_{\rm max} = (8/3)^\frac25 a_{\rm end}$ \cite{gkmo1,cmov} and the radiation density prior to reheating evolves as 
\beq
\rho_{\rm R}(a) = \frac{2\sqrt{3}}{5}\Gamma_\phi M_P \sqrt{\rhoend} \left( \frac{\aend}{a} \right)^4 \left[ \left( \frac{a}{\aend} \right)^\frac52 - 1 \right] \, ,
\label{eq:rhoR}
\eeq
for $\aend < a < \arh$
where for now we have ignored any possible contribution to $\rho_{\rm R}$ from $\chi$ decays. 

We define the reheating temperature as the temperature of the radiation bath (produced by inflaton decays) when the energy density of the radiation and the energy density left in the inflaton condensate are equal at $a = \arh$ , i.e. $\rho_{\phi}(a_{\rm RH})=\rho_{R}(a_{\rm RH})$. Using (\ref{eq:rhophi}) and (\ref{eq:rhoR}), one finds
\beq
\left( \frac{\arh}{\aend} \right)^\frac32 \simeq \frac{5\sqrt{\rhoend}}{2\sqrt{3} \Gamma_\phi M_P} \, ,
\eeq
and therefore
\beq
\rhorh = \rho_{\phi}(a_{\rm RH}) = \rhoend \left( \frac{\aend}{\arh} \right)^3 = \frac{12}{25} \left( \Gamma_\phi M_P \right)^2 \equiv \alpha \trh^4\,,
\label{rhorh}
\eeq
where $\alpha = g_{\rm RH}\pi^2/30$ and $g_{\rm RH} = 427/4$ is the number of Standard Model degrees of freedom at sufficiently high temperature. For $a > \arh$,
we can write
\beq
\rho_{\rm R}(a) = \rhorh \left( \frac{\arh}{a} \right)^4\,.
\eeq

Finally, for $a>a_{\chi}$, the energy density in $\chi$ is given by 
\be
\rho_{\chi}(a) = \frac12 m_{\chi}^2 \chi_{\rm inf}^2 \left( \frac{a_\chi}{a} \right)^3 = \frac{3}{16 \pi^2} H_{\rm end}^4 \left( \frac{a_\chi}{a} \right)^3 \, ,
\label{eq:rhochi}
\ee
where the second equality comes from (\ref{eq:sigmainf}). For $\aend < a < a_\chi$, $\rho_{\chi}$ remains constant with $\rho_\chi = \frac{3}{16 \pi^2} H_{\rm end}^4$. $\chi$ oscillations begin when at $a = a_\chi$ defined by $\frac32 H = m_{\chi}$ when the total energy density is dominated by matter and by $2 H = \mchi$ when the energy density is dominated by radiation. The determination of $a_\chi$ will thus depend on whether or not the inflaton has decayed. Similarly, $\chi$ decays occur when $a = a_{d\chi}$ defined by
$\frac32 H = \Gamma_\chi$ or $2H = \Gamma_\chi$ for matter and radiation domination (MD and RD) respectively.\footnote{The coefficient of $H$ relative to $\mchi$ or $\Gamma_\chi$ depends on the ratio of the radiation to matter density, $z$, 
which tends to $\frac32$ and 2 at large and small $z$, respectively. See Appendix~B for the functional form of this coefficient.
} Like  $a_\chi$, the determination of $a_{d\chi}$ will depend on $\Gamma_\phi$. 

Given the above, there are three logical possibilities:
\begin{enumerate}
    \item $\aend < a_\chi < a_{d\chi} < \arh$
    \item $\aend < a_\chi  < \arh < a_{d\chi} $
    \item $\aend   < \arh < a_\chi < a_{d\chi} $
\end{enumerate}
In the first case, though we will derive the ratios $a_\chi/\aend$ and $a_{d\chi}/\aend$ for completeness, this case is not particularly interesting as $\chi$ decays as a sub-dominant species before reheating. In order to avoid a second period of inflation we enforce $\rho_{\chi}(a_{\chi})<\rho_{\phi}(a_{\chi})$. When $\chi$ starts to oscillate, the energy of the universe is still dominated by the inflaton, and the ratio $\rho_{\phi}/\rho_{\chi}$ remains constant until $a=a_{d\chi}$. For $a_{d\chi}<a<a_{\rm RH}$, the ratio $\rho_{r\chi}/\rho_{\phi}$ decreases until $a_{\rm RH}$ with $\rho_{r\chi}$ the energy density of radiation from $\chi$-decay. This means the matter domination era continues for $a_{\rm end}<a<a_{\rm RH}$ for case 1. As $\chi$ or its decay products never come to dominate the energy of universe until $a_{\rm RH}$, $\chi$ cannot make a significant contribution to $n_{s}$ and $r$.
In the latter two cases, we will further distinguish whether or not $\chi$ decays during a matter ($\chi$-dominated) or radiation dominated era.

\section{Scenarios}
\label{sec:scenarios}

In this section, we will consider 
the three evolutionary scenarios classified by different values of three free parameters
\be
\text{Free parameters}:\quad \mchi,\,\,\Gamma_\chi,\,\,\trh\,.
\ee  
Every choice of $(\mchi, \Gamma_\chi, \trh)$ will correspond to a specific evolutionary scenario and thus to unique predictions for the inflationary observables. 

\subsection{Case I}

As noted earlier, bearing in mind standard cold inflationary models, we assume that the Universe is dominated by the inflaton when $\chi$ oscillations begin. Then, the expansion rate for $a_{\rm end}<a<a_{\chi}$ is given by $H(a)^2 = \rho_\phi(a)/3M_P^2$. Oscillations of $\chi$ begin when $\frac32 H = m_{\chi}$. Similarly $\chi$ decays occur when $\frac32 H = \Gamma_\chi$. From these, the scale factors when $\chi$ oscillations begin and when $\chi$ decays are easily obtained:
\be
\left( \frac{a_\chi}{\aend} \right)^3 = \frac34 \frac{\rhoend}{m_{\chi}^2 M_P^2}\,,
\label{eq:acae1}
\ee
and
\be
\left( \frac{a_{d\chi}}{\aend} \right)^3 = \frac34 \frac{\rhoend}{\Gamma_\chi^2 M_P^2} \, .
\label{eq:adcae1}
\ee

Demanding the absence of the second inflationary phase driven by $\chi$ will require that the energy density of $\chi$ does not dominate the total energy before $\chi$ begins its oscillations. From (\ref{eq:rhophi}), (\ref{eq:rhochi}) and (\ref{eq:acae1}), this leads to 
\be
\frac{3}{16 \pi^2} H_{\rm end}^{4}< \frac43 m_{\chi}^{2}M_{P}^{2}<\rho_{\rm end}\,.
\label{basic}
\ee
 The first of these inequalities comes from $\rho_{\chi}(a_{\chi})<\rho_{\phi}(a_{\chi})$.
 The second inequality in Eq.~(\ref{basic}) is then derived simply from $a_{\rm end}<a_{\chi}$.
A violation of this inequality, namely $m_{\chi} < 3 H_{\rm end}^2/8 \pi M_P$, would lead to a second period of inflation driven by $\chi$.  
Clearly we must have $H_{\rm end}^4 < 16 \pi^2 \rhoend/3 $ to allow a finite mass range in $m_{\chi}$\footnote{Throughout, by $m_\chi$, we mean the effective mass of $m_{\chi}$. In the absence of self-interactions or couplings to the inflaton $m_{\chi,\rm eff}= m_{\chi}$. Including these interactions would increase $m_{\chi,\rm eff}$ and therefore decrease $\langle \chi_{\rm inf}^2 \rangle$ and $\rho_\chi$. Thus, bare masses in violation of the left side of the inequality in (\ref{basic}) simply require interaction which force the effective mass to satisfy (\ref{basic}). }.  

The energy density of $\chi$ at the time of decay is 
\begin{align}
\rho_\chi|_{d\chi}  =  \frac12 m_{\chi}^2 \chi_{\rm inf}^2 \left( \frac{a_\chi}{a_{d\chi}} \right)^3  & =  \frac{3}{16\pi^2} H_{\rm end}^4 \left( \frac{a_\chi}{a_{d\chi}} \right)^3 \nonumber \\
& = \frac{3}{16\pi^2} H_{\rm end}^4 \frac{\Gamma_\chi^2}{\mchi^2}\,,
\end{align}
which can be compared with the energy density in inflaton oscillations at the same time
\be
\rho_\phi|_{d\chi} = \rhoend \left( \frac{\aend}{a_{d\chi}} \right)^3 = \frac43 \Gamma_\chi^2 M_P^2\,.
\ee
Since $a_\chi < a_{d\chi}$, we must have $\Gamma_\chi < m_\chi$ and for $a_{d\chi} < \arh$, we must have $
\Gamma_{\chi} > \sqrt{3\alpha} \trh^2/2M_P$. 
Putting these two constraints together,
this case must satisfy
\be
\frac{5 \sqrt{\alpha}\trh^2}{\sqrt{12} M_P} = \frac35 \Gamma_\phi < \Gamma_\chi < m_{\chi} \, .
\ee
However, as noted earlier, since neither $\chi$ nor its decay products provide a significant contribution to the total energy $(\omega_\chi \ll 1)$, this case will not be able to affect the inflationary observables.

\subsection{Case II}
In this case, the ratio $a_\chi/\aend$ is again given by Eq.~(\ref{eq:acae1}). When $\chi$ decays, the universe could be in either a RD era or a MD era.

\underline{{\bf Case II-RD}}: If $\chi$ decays occur in a RD background 
with $H(a>a_{\rm RH})^2 = (\rhorh/3 M_P^2) (\arh/a)^4 $ one finds
\be
\left( \frac{a_{d\chi}}{\aend} \right)^3 = \left( \frac43 \right)^\frac34 \frac{\rhoend}{\alpha^\frac14 \trh (\Gamma_\chi M_P)^\frac32} \, .
\label{adxae2a}
\ee
At $a=a_{d\chi}$, the energy density in $\chi$ is
\begin{align}
\rho_\chi|_{d\chi}  & = \frac12 m_{\chi}^2 \chi_{\rm inf}^2 \left( \frac{a_\chi}{a_{d\chi}} \right)^3  = \frac{3}{16\pi^2} H_{\rm end}^4 \left( \frac{a_\chi}{a_{d\chi}} \right)^3 \nonumber \\
& = \frac{3}{16\pi^2} \left(\frac34 \right)^\frac74 H_{\rm end}^4 \frac{\alpha^\frac14 \trh \Gamma_\chi^\frac32}{m_{\chi}^2 M_P^\frac12} \, ,
\label{rhodchi2a}
\end{align}
which can be compared to the energy density in radiation at the same time
\be
\rho_r|_{d\chi} = \rhorh \left( \frac{\arh}{a_{d\chi}} \right)^4 = \frac34 \Gamma_\chi^2 M_P^2 \, .
\label{rhordc2}
\ee
From Eq.~(\ref{rhordc2}), we also see that $\chi$ decays occur before big bang nucleosynthesis (BBN) if $\Gamma_\chi \gtrsim 10^{-22}$~GeV.

To compute $R_\chi$, from (\ref{rhodchi2a}) and (\ref{rhordc2}) we first compute 
\be
x = \frac{\rho_{\chi}}{\rho_{r}}\biggr\rvert_{d\chi}=\frac{3}{16\pi^2} \left(\frac34 \right)^\frac34 \frac{H_{\rm end}^4 \alpha^\frac14 \trh}{m_{\chi}^2 \Gamma_\chi^\frac12 M_P^\frac52} \, .
\label{x2a}
\ee
These results hold so long as 
\be
\text{(i)}:\,\,\,\frac34 \Gamma_\chi M_P < \frac12 \sqrt{3 \alpha} \trh^2 < m_{\chi} M_P < \frac12 \sqrt{3 \rhoend} \, ,
\label{eq:caseIIahierarchy}
\ee
from $\aend < a_\chi  < \arh < a_{d\chi} $ and
\be
\text{(ii-\text{RD})}:\,\,\,\Gamma_\chi > \frac{27\sqrt{3}}{2048     \pi^4} \frac{H_{\rm end}^8\sqrt{\alpha}\trh^2}{m_{\chi}^4 M_P^5} \, .
\label{gammalow}
\ee
from \text{$\rho_{\chi}(a_{d\chi})<\rho_{r}(a_{d\chi})$}, i.e. $x<1$.

\underline{{\bf Case II-MD}}: For sufficiently small $\Gamma_\chi$, $\chi$-oscillations will come to dominate over the radiation from inflaton decays, leading to a second period of matter domination. 
Matter radiation equality occurs at $a_{\rm eq}$ when
\begin{align}
\rho_\chi(a_{\rm eq}) = & \frac{3}{16\pi^2} H_{\rm end}^4 \left( \frac{a_\chi}{a_{\rm eq}} \right)^3 \nonumber \\
 =  & \alpha \trh^4 \left( \frac{\arh}{a_{\rm eq}} \right)^4 = \rho_r(a_{\rm eq}) \, ,
\end{align}
leading to 
\be
\frac{\arh}{a_{\rm eq}} = \frac{9}{64\pi^2} \frac{H_{\rm end}^4}{m_{\chi}^2 M_P^2} \, .
\ee
When $\chi$ decays in a MD period, we must recompute $a_{d\chi}$ and we obtain
\be
\left( \frac{a_{d\chi}}{\aend} \right)^3 = \frac{81}{256\pi^2} \frac{ H_{\rm end}^6}{m_{\chi}^2 \Gamma_\chi^2 M_P^2} \, .
\label{adcae2b}
\ee
Then $a_{d\chi}>a_{\rm eq}$ is equivalent to 
\be
\text{(ii-\text{MD})}:\,\,\,\Gamma_\chi < \frac{81\sqrt{3}}{8192 \pi^4} \frac{H_{\rm end}^8\sqrt{\alpha}\trh^2}{m_{\chi}^4 M_P^5} \, .
\label{gammachi2b}
\ee
Recomputing $x$, we find
\be
x = \left( \frac34 \left(\frac{9}{64\pi^2}\right)^4 \frac{\alpha \trh^4 H_{\rm end}^{16}}{m_{\chi}^8 \Gamma_\chi^2 M_P^{10}} \right)^\frac13 \, .
\label{x2b}
\ee

When $\chi$ decays in a MD era, there is entropy production quantified by $s_f/s_i = x^\frac34$ and a new reheating temperature determined by $\Gamma_\chi$. For successful BBN, we again require $\Gamma_\chi \gtrsim 10^{-22}$~GeV. The amount of entropy production depends on $\trh$ and $H_{\rm end}$ and $\mchi$ through $x$ in (\ref{x2b}).

In sum, (\ref{eq:caseIIahierarchy}) applies to both of the cases II-RD and II-MD. The two cases differ in the evaluation of $x$ in (\ref{x2a}) and (\ref{x2b}) and the conditions  on $\Gamma_\chi$ (\ref{gammalow}) and (\ref{gammachi2b}).\footnote{More accurately, as $a_{d\chi} \to a_{\rm eq}$, the expressions  should be subject to the rescalings $\Gamma_\chi\rightarrow 2\Gamma_\chi/\beta_x$ and $\Gamma_\chi\rightarrow 3\Gamma_\chi/2\beta_x$, for radiation and matter dominated eras respectively, where $\beta_x = \frac34(\sqrt{2}+1)$. This is because at $a_{\rm eq}$, $Ht= 1/\beta$ rather than (1/2) or (2/3). Similarly, as $a_\chi\rightarrow a_{\rm RH}$, the expressions in case II and case III should be rescaled by $m_\chi\rightarrow 3m_\chi/2\beta_y $ and $m_\chi \rightarrow 2m_\chi/\beta_y$. Away from $a_{d\chi}=a_{\text{eq}}$ or $a_\chi=a_{\text{RH}}$,   the same rescalings apply, with the functional form of $\beta_x$ and $\beta_y$ given in Appendix B.
}

\subsection{Case III}
Case III differs from Case II in that reheating occurs earlier than the onset of $\chi$-oscillations, i.e. when $a_\chi > \arh$ or when $\frac45 \Gamma_{\phi}>m_{\chi}$. To determine when oscillations begin, we must use the expression of Hubble expansion rate in a RD era and we find
\be
\left( \frac{a_\chi}{\arh} \right)^4 = \frac43 \frac{\alpha \trh^4}{m_{\chi}^2 M_P^2} \,.
\label{acrh3a}
\ee

\underline{{\bf Case III-RD}}:  If the universe is in a RD era when $\chi$ decays
\be
\left( \frac{a_{d\chi}}{\arh} \right)^4 = \frac43 \frac{\alpha \trh^4}{\Gamma_\chi^2 M_P^2} \,.
\label{acrhd3a}
\ee
At $a_{d\chi}$, we have
\be
\rho_\chi(a_{d\chi}) =  \frac{3}{16\pi^2} H_{\rm end}^4 \left( \frac{a_\chi}{a_{d\chi}} \right)^3 = \frac{3}{16\pi^2} H_{\rm end}^4 \left(\frac{\Gamma_\chi}{m_{\chi}} \right)^\frac32\,.
\label{rhochi3a}
\ee
By comparing (\ref{rhochi3a}) to $\rho_{r}(a_{d\chi})$ which is still given by Eq.~(\ref{rhordc2}), one obtains
\be
x = \frac{1}{4\pi^2} \frac{H_{\rm end}^4}{m_{\chi}^\frac32 \Gamma_\chi^\frac12 M_P^2} \, .
\label{x3a}
\ee
These result hold so long as 
\be
\text{(i)}:\,\,\,\Gamma_\chi  < m_{\chi} < \frac{2}{\sqrt{3}}   \frac{\sqrt{\alpha} \trh^2}{M_P}< \frac{2}{\sqrt{3}}   \frac{\sqrt{\rho_{\rm end}}}{M_P}\, .
\label{eq:caseIIIRDhierarchy}
\ee
from the defining condition for the case III , i.e. $\aend   < \arh < a_\chi < a_{d\chi} $ and
\be
\text{(ii-RD)}:\,\,\,\frac{H_{\rm end}^{8}}{16\pi^{4}m_{\chi}^{3}M_{P}^{4}}<\Gamma_{\chi}\,,
\label{eq:gammalowIII}
\ee
from \text{$\rho_{\chi}(a_{d\chi})<\rho_{r}(a_{d\chi})$}, i.e. $x<1$.\footnote{The same condition can be obtained from $a_{\rm eq}>a_{d\chi}$.} As in Case II, $\chi$ decays before BBN imply $\Gamma_\chi \gtrsim 10^{-22}$~GeV.

\underline{{\bf Case III-MD}}: Once again, for sufficiently small $\Gamma_{\chi}$ (late $\chi$ decay), it is also possible for $\chi$ to decay in a MD era. This occurs if $a_{\chi}<a_{\rm eq}<a_{d\chi}$ is satisfied. 
From (\ref{acrh3a}) and $\rho_{R}(a_{\rm eq})=\rho_{\chi}(a_{\rm eq})$, one can find
\be
\frac{\arh}{a_{\rm eq}} = \frac{3}{16\pi^2} \left( \frac43 \right)^\frac34 \frac{H_{\rm end}^4}{\alpha^\frac14 \trh m_{\chi}^\frac32 M_P^\frac32} \, .
\ee
Recomputing $a_{d\chi}$ from (\ref{acrh3a}) and $(3/2)H(a_{d\chi})=\Gamma_{\chi}$, one finds
\be
\left( \frac{a_{d\chi}}{\arh} \right)^3 = \frac{3}{16\pi^2} \left( \frac34 \right)^\frac14 \frac{H_{\rm end}^4 \alpha^\frac34 \trh^3}{m_{\chi}^\frac32 \Gamma_\chi^2 M_P^\frac72}\,.
\label{achrhd3b}
\ee
The condition
$a_{\rm eq}<a_{d\chi}$ is equivalent to 
\be
\text{(ii-MD)}:\,\,\,\Gamma_{\chi}<\frac{3H_{\rm end}^{8}}{64\pi^{4}m_{\chi}^{3}M_{P}^{4}}\,.
\label{eq:gammalowIIIMD}
\ee
Then, the comparison of $\rho_{R}(a_{d\chi})$ to $\rho_{\chi}(a_{d\chi})$ yields
\be
x =  \left( \frac{9}{4096\pi^8} \frac{H_{\rm end}^{16}}{
m_{\chi}^6 \Gamma_\chi^2 M_P^8
} \right)^\frac13 \,.
\label{x3b}
\ee
As in case II-MD, there is entropy production and a new reheating temperature. 

In sum, (\ref{eq:caseIIIRDhierarchy}) applies to both of the cases III-RD and III-MD. Two cases differ in the evaluation of $x$ in (\ref{x3a}) and (\ref{x3b}), and the conditions (\ref{eq:gammalowIII}) and (\ref{eq:gammalowIIIMD}). Notice that expressions of $x$ of both cases are independent of $\trh$.

\section{Application to Models}
\label{sec:app}
The discussion in the previous section is model-independent. As we discussed above,
the parameter space under consideration is described by $\trh$, $\mchi$ and $\Gamma_\chi$. Once specified, each set of these three parameters will correspond to a model with an evolution described by cases I, II, or III (and whether or not curvaton decay occurs in a RD or MD universe). However, the precise boundaries of these cases also depends on $H_{\rm end}$ and therefore is not completely model independent. 

Furthermore, the derived constraints on the curvaton scenario, in particular its mass range and decay rate, depend on the specific inflation model. While $R_\chi$ and $\omega_\chi$ only carry the dependence on $H_{\rm end}$, the observables $n_s$ and $r$ depend on the inflation slow-roll parameters, and even the differences $\delta n_s$ and $\delta r = -r_* \omega_\chi$ depend explicitly on $\epsilon_*$ and $\eta_*$.

Here, we illustrate the curvaton constraints in two well-motivated examples, the first is the  Starobinsky model \cite{Staro} which is in good agreement with CMB bounds, and the second is a new inflation model \cite{new}, which predicts a value of $n_s$ significantly below the measured value.  In the former model, the curvaton is constrained so that it does not upset the already good agreement with experiment. Nevertheless, for certain parameters the already good agreement is actually improved. For the latter, the contribution from the curvaton becomes important, as it is capable of raising the prediction in $n_s$ up to the experimental value.

Both of the models we consider can be derived in the context of no-scale supergravity \cite{no-scale} as described in Appendix A.  
In the new inflation model, 
the inflaton evolves from a very small field value ($\phi \ll M_P$) to a minimum at $\phi \approx M_P$. 
The scalar potential in this case is
\begin{align}
    V_n(z)  = 
    & M^2 M_P^2 \left( 1-6\sqrt{6}\sinh \left(\frac{z}{\sqrt{6}}\right)^2\tanh \left(\frac{z}{\sqrt{6}}\right)\right)^2 
    \label{newpotential}\\
     \simeq & M^2 M_P^2 (z^3-1)^2 \label{eenospot} \, ,
\end{align}
where $z=\phi/M_P$ is the canonical inflaton field value normalized by the Planck mass. The inflaton mass for this potential is $m_\phi = \sqrt{18} M$.
For $N_*=55$, the new inflation model gives an extremely small tensor-to-scalar ratio $r\simeq 2.4\times 10^{-8}$, whereas the scalar spectral index is $n_s\simeq 0.928$, significantly lower than the CMB determination. For this reason, this type of model of inflation was considered excluded.

In the Starobinsky model,
the inflaton evolves from a plateau at large field values to a minimum at $\phi = 0$.
The potential in this case is
\be
V_S(z)=\frac{3}{4} M^2 M_P^2  \left(1-e^{-\sqrt{\frac23}z}\right)^2\, ,
\label{Spotential0}
\ee
where again $z=\phi/M_{P}$ is the inflaton normalized by the Planck scale. In this case, the inflaton mass is $m_\phi = M$.
For $N_*=55$, we have $r\simeq 3.5\times 10^{-3}$, and $n_s\simeq 0.965$, both in good agreement with CMB observations.

 To see the dependence of the cosmological observables on the reheating temperature, recall that the number of $e$-folds can be computed by \cite{LiddleLeach,MRcmb}:
\begin{align}
N_* & =\ln \left[\frac{1}{\sqrt{3}}\left(\frac{\pi^2}{30}\right)^{1 / 4}\left(\frac{43}{11}\right)^{1 / 3} \frac{T_0}{H_0}\right]-\ln \left(\frac{k_*}{a_0 H_0}\right)  \nonumber \\
&-\frac{1}{12} \ln g_{\text {RH}} +\frac{1}{4} \ln \left(\frac{V_*^2}{M_P^4 \rho_{\text {end }}}\right)+\ln R_\text{rad}\,,
\label{Nstar}
\end{align}
where $H_0=67.36 \mathrm{~km} \mathrm{~s}^{-1} \mathrm{Mpc}^{-1}$ is the present day Hubble parameter, and we take $a_0=1$. The last term above is the reheating parameter, that we approximate as
\beq 
\ln R_\text{rad}\simeq \frac{1}{6}\ln \left(\frac{\Gamma_\phi}{H_\text{end}}\right)\,.
\label{Rrad}
\eeq
The reheating temperature is related to the inflaton decay rate through Eq.~(\ref{rhorh}).
For concreteness, we consider reheating through a Yukawa-like coupling $y$, so the inflaton decay rate is given by:
\begin{equation}\Gamma_\phi=\frac{y_\phi^2}{8\pi }m_\phi \, ,
\end{equation}
where $m_\phi$ is the inflaton mass.

\begin{figure}[ht!]
  \centering
\includegraphics[width=0.45\textwidth]{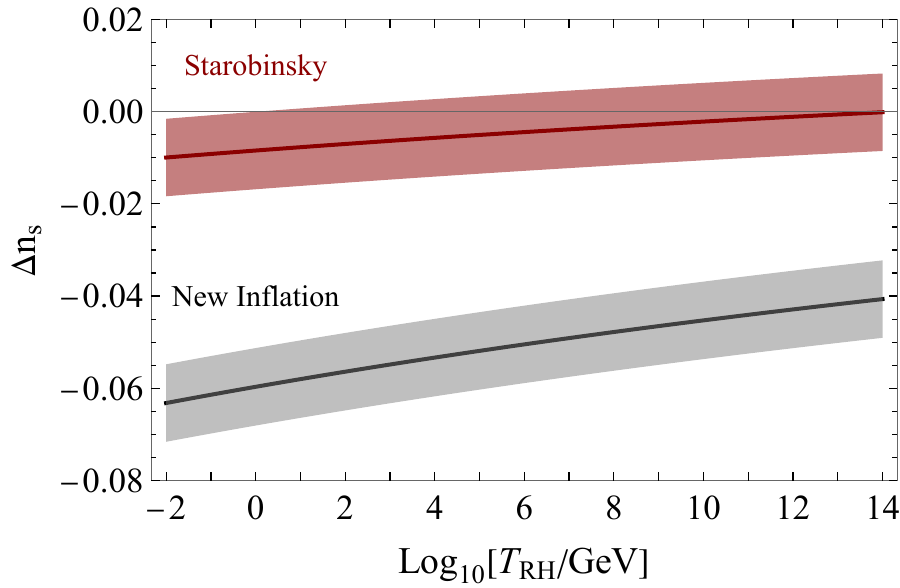}
  \caption{\em \small {The difference, $\Delta n_s=n_*-n_{s,\rm{CMB}}$,  between the model prediction of $n_s$ from new inflation and the Starobinsky model, and the value determined from CMB observations \cite{Planck}. We take $n_{s,\rm{CMB}}=0.9649$ (solid lines) with $\pm 2\sigma=0.0084$ standard deviations (shaded regions). 
} 
}
  \label{nsprediction}
\end{figure}

Insisting on a perturbative coupling forbids very high values of $\trh$. The prediction of $n_s$ in a given model is obtained by \eqref{ns}, and its dependence on $\trh$ enters through the determination of $N_*$ (through Eqs.~(\ref{Nstar}) and (\ref{Rrad})). In Fig.~\ref{nsprediction}, we show as a function of $T_{\rm RH}$ the deviation $\Delta n_s$ of the predicted $n_{s}$ of Starobinsky model and new inflation model without introducing the curvaton from the observed $n_{s,{\rm CMB}}$ from the CMB temperature anisotropy power spectrum \cite{Planck}. The solid lines show the difference between $n_{s}$ and the Planck central value of $n_{s,{\rm CMB}} = 0.9649$ and the band corresponds to the difference between $n_{s}$ and the 2$\sigma = 0.0084$ range.
For all $T_{\rm RH}$ below the Planck scale, the new inflation model prediction of $n_{s}$ is always below experimental constraints $0.958<n_s<0.975$, with the discrepancy being of order $0.05$. In contrast, the Starobinsky model alone can successfully account for the observed $n_{s}$ within $2\sigma$ for any $\trh \gtrsim 4$~GeV, and for sufficiently high $T_{\rm RH}$, even the central Planck value can be attained. Therefore, if the inflaton sector of these models is extended by including the curvaton, from (\ref{eq:deltans}), we expect that the new inflation model requires a much more weakly coupled curvaton (smaller $\Gamma_{\chi}$) than the Starobinsky model for a fixed $T_{\rm RH}$ and $m_{\chi}$ since smaller $\Gamma_{\chi}$ corresponds to a later decay of $\chi$ and is expected to yield a larger value of $R_{\chi}$.

From Eq.~(\ref{eq:deltans}), we can compute the change in the tilt, $\delta n_s$ and from Eq.~(\ref{deltar}), the change in the tensor-to-scalar ratio, $\delta r$. Notice first that the shift in $r$ is always toward lower $r$. The shift in $n_s$ is more model dependent. However, 
$\eta_\chi = M_{P}^{2}(\mchi^2/V)$.
For the Starobinsky model, this gives $\eta_\chi = \frac43 \mchi^2/m_\phi^2$ and for the new inflation model, $\eta_\chi = 18 \mchi^2/m_\phi^2$. In both cases, the contribution is positive and presumably small (for $\mchi \ll m_\phi$). For $N_* = 55$, 
$\epsilon_* \simeq 0.00022$ and $\eta_* = -.0169$ for the Starobinsky model and $\epsilon_* \simeq 1.5 \times 10^{-9}$ and $\eta_* = -.036$ for the new inflation model. In both cases, $\delta n_s$ is always positive and it may be possible to find a set of parameters such that $\delta n_s + \Delta n_s = n_{s}-n_{s,{\rm CMB}} = 0$.

\begin{figure*}[ht!]
\centering
\hspace*{-5mm}
\subfigure{\includegraphics[width=0.45\textwidth]{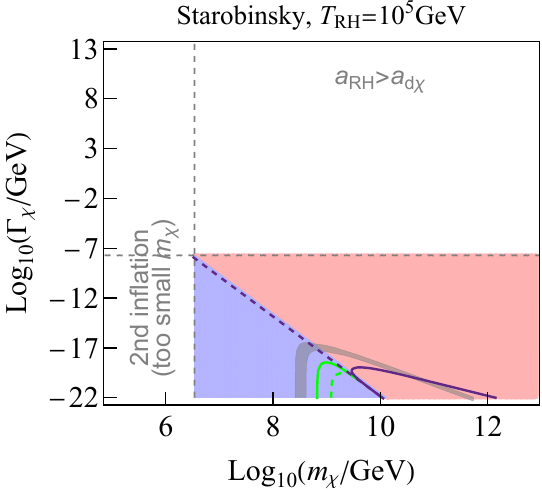}}\qquad
\subfigure{\includegraphics[width=0.45\textwidth]{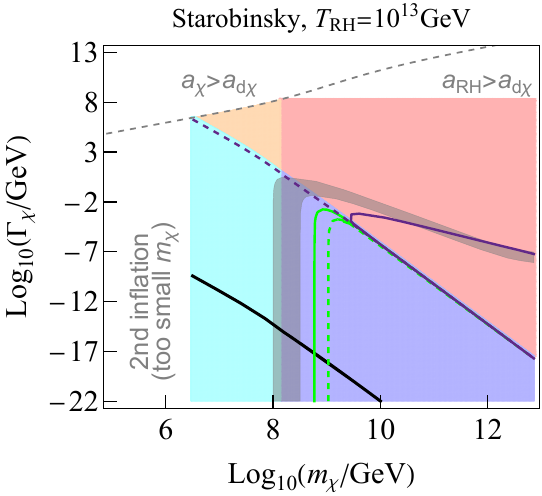}}\\
\subfigure{\includegraphics[width=0.45\textwidth]{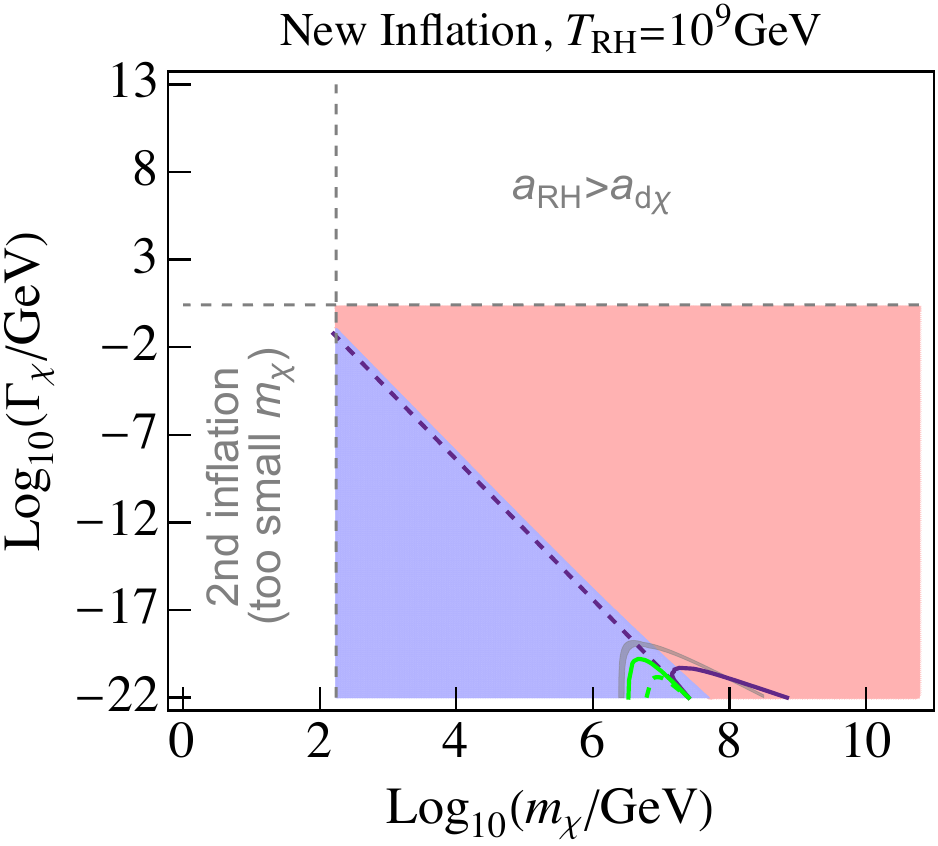}}\qquad
\subfigure{\includegraphics[width=0.45\textwidth]{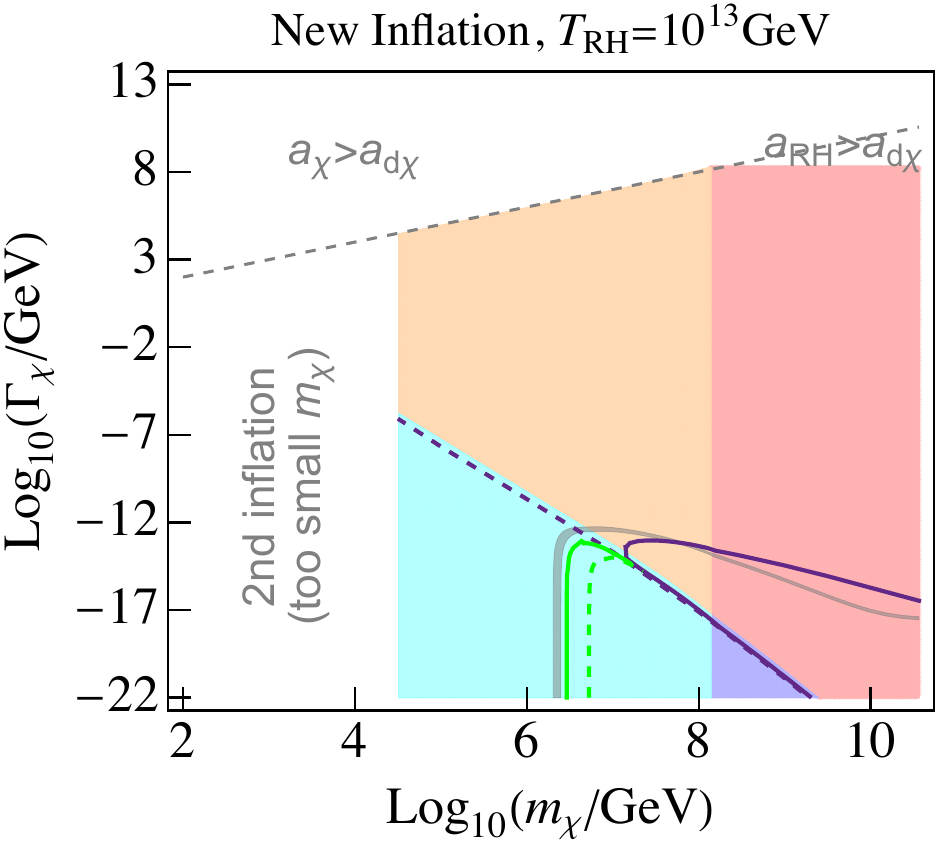}}
\caption{The $m_{\chi}$ vs $\Gamma_{\chi}$ parameter space for a curvaton in the Starobinsky model (upper panels) and the new inflation model (lower panels) for $T_{\rm RH}=10^{5}, 10^{9}{\rm GeV}$ (left panels) and $T_{\rm RH}=10^{13}{\rm GeV}$ (right panel). The background shading corresponds to cases II-RD (red), II-MD (blue), III-RD (yellow) and III-MD (cyan). The unshaded regions correspond to an exclusion due to a second period of inflation (low $\mchi$), case I (high $\Gamma_\chi$), or $a_{d\chi} < a_\chi$ (also high $\Gamma_\chi$ at high $\trh$).  Within the gray bands shown for the Starobinsky model, $\Delta n_{s}+\delta n_s = 0$ including $-2\sigma$ deviations.
All of the plane to the left and above this band is acceptable within $+2\sigma$.
For the new inflation model, the band corresponds to $\Delta n_{s}+\delta n_s = 0$ with $\pm 2\sigma$ deviations. 
The set of parameters $m_{\chi}$, $\Gamma_{\chi}$, $T_{\rm RH}$ along the purple (green) solid and dashed lines yields $f_{NL}=9$ and $f_{NL}=0$ ($f_{NL}=-1$ and $f_{NL}=-11$) respectively.
The black line seen in the upper right panel corresponds to $x= 5\times 10^{10}$, below which there is an excess of entropy production when $\chi$ decays.
}
\vspace*{-1.5mm}
\label{fig2}
\end{figure*}

Given the classifications of the different cosmological scenarios from $a_{\rm end}$ to $a_{d\chi}$ in Sec.~\ref{sec:scenarios}, for each $T_{\rm RH}$, one can specify a pair of curvaton parameters $(m_{\chi},\Gamma_{\chi})$  that enables the net $n_{s}$ in (\ref{eq:netns}) to perfectly fit the observed scalar spectral index from the CMB. In Fig.~\ref{fig2}, for two specific reheating temperatures in each of the two inflationary models considered, we show the result of probing the $(m_{\chi},\Gamma_{\chi})$ parameter space in which the curvaton provides a solution to $\delta n_s + \Delta n_s = 0$, that is where the curvaton contribution to $n_s$ can bridge a deficit in the model. 
Recall that since $\delta n_s$ is always positive, such a solution only exists if $\Delta n_s < 0$.

The upper two panels in Fig.~\ref{fig2} show results for the Starobinsky model for fixed $T_{\rm RH}=10^{5}$~GeV (left) and  $10^{13}~{\rm GeV}$ (right). The lower panels show the results for the new inflation model with $T_{\rm RH}=10^{9}$~GeV (left) and  $10^{13}~{\rm GeV}$ (right). The background shaded regions distinguish case II-RD (blue), II-MD (red), III-RD (cyan) and III-MD (yellow). Case I corresponding to $a_{d\chi} < \arh$ is found in the unshaded region at high $\Gamma_\chi$. 
The unshaded regions at small $\mchi$
are not viable because a second period of inflation ensues. We also delineate the region where $\chi$ decays before the onset of oscillations ($a_{\chi}>a_{d\chi}$).

As discussed earlier, we take $\Gamma_{\chi}\gtrsim 10^{-22}~{\rm GeV}$ to insure that $\chi$-decay in a RD era occurs before the temperature of the primordial plasma reaches $T\simeq~10~{\rm MeV}$ and that the new reheating temperature is sufficiently high when $\chi$ decays in a MD era. This guarantees that light element production during BBN is not spoiled by late time $\chi$-decay. Since case III requires early inflaton reheating (to ensure $a_{\rm RH} < a_\chi)$ this case is only realized for sufficiently high $T_{\rm RH}$. This explains why the yellow and cyan shaded background appear only for higher $T_{\rm RH}$ (seen in the right panels). 

We also noted earlier, that when $\chi$ decays and dominates the energy density,
entropy is produced during this second period of reheating. The amount of entropy production is determined by the value of $x$, that is the ratio $\rho_\chi/\rho_r$ and $s_f/s_i = x^\frac34$. For very large $x$ even if BBN proceeds normally, a large amount of entropy production can over-dilute any existing baryon asymmetry. If we
assume a maximal asymmetry of $n_B/s \sim 10^{-2}$, then the largest allowed value of $x$ is $\sim 5 \times 10^{10}$. This constant value of $x$ appears as a thick black diagonal line and is shown on the upper right panel in Fig.~\ref{fig2}.
All points below this line are excluded as the entropy production becomes excessive. For the new inflation model, $x<5 \times 10^{10}$ holds for all of the displayed blue and cyan-shaded region corresponding to II,III-MD.

\begin{figure}[ht!]
  \centering
\includegraphics[width=0.45\textwidth]{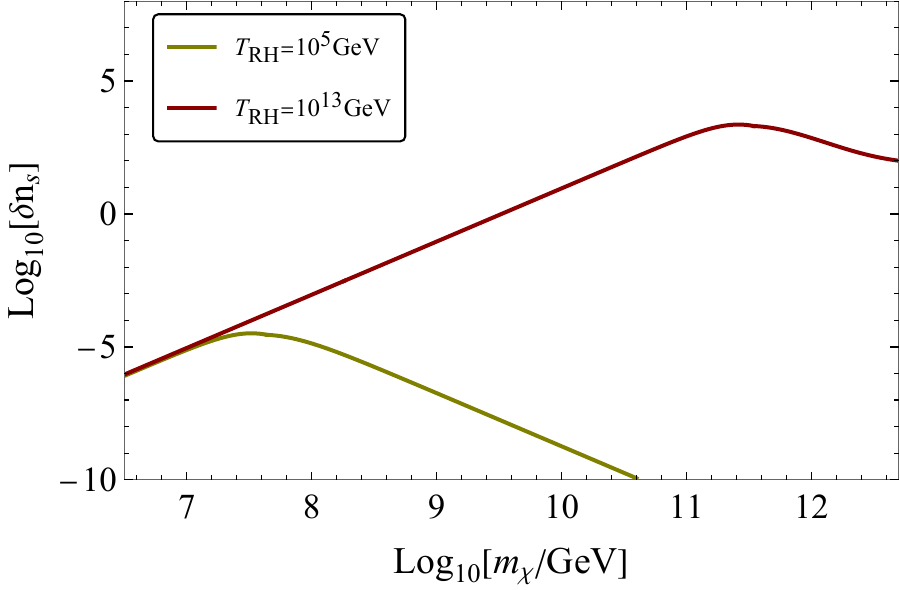}
  \caption{\em \small $\delta n_{s}$ as a function of $m_\chi$ for the Starobinsky model, with fixed $\Gamma_\chi=10^{-12}~{\rm  GeV}$ for $\trh = 10^5$~GeV (red) and $10^{13}$~GeV (green).
} 
  \label{dnsmchi}
\end{figure}

\begin{figure}[ht!]
  \centering
\includegraphics[width=0.45\textwidth]{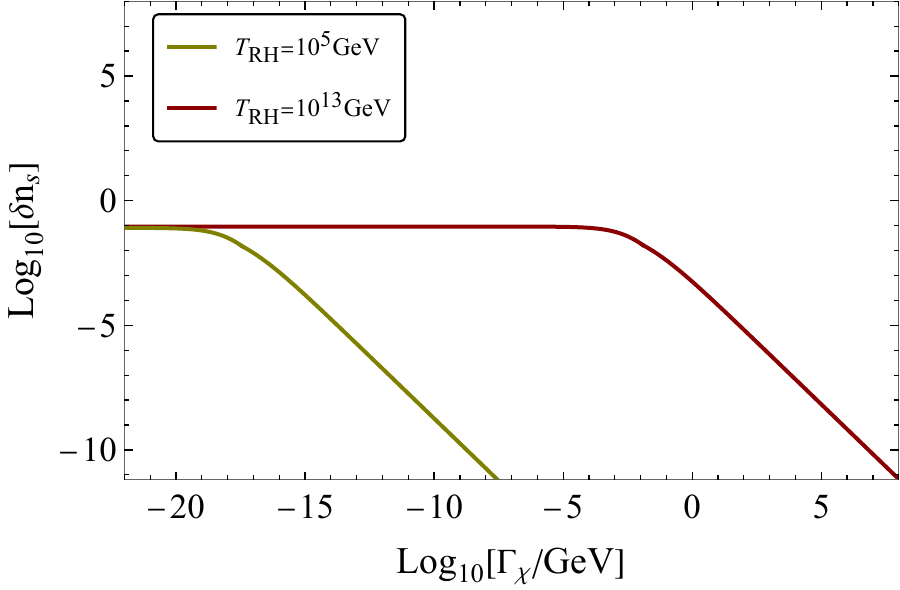}
  \caption{\em \small {$\delta n_{s}$ as a function of $\Gamma_\chi$ for the Starobinsky model, with fixed  $m_\chi=10^{9}~{\rm  GeV}$ for $\trh = 10^5$~GeV (red) and $10^{13}$~GeV (green).
} 
}
  \label{dnsgamma}
\end{figure}

The regions in the $(\mchi,\Gamma_\chi)$
plane with acceptable values of $n_s$ are shown by the gray bands in Fig.~\ref{fig2}.
However the meaning of the two bands is different for the Starobinsky and new inflation models. For the Starobinsky
model, the left/upper side of the band corresponds to $\Delta n_s + \delta n_s =0$
whereas the right/lower edge of the band
corresponds to $\Delta n_s + \delta n_s = 0.0084$ corresponding to the 2$\sigma$ lower bound on the experimental value of $n_s$. Since the Starobinsky model already matches the observation quite well, the model is within the 2$\sigma$ upper bound for all points above and to the left of the gray band. Only the points in the lower left corner of the plane are excluded as they provide a correction $\delta n_s$ in excess of the predicted value of $\Delta n_s$ shown in Fig.~\ref{nsprediction}. This can perhaps be better understood from Fig.~\ref{dnsmchi} which shows the behavior of $\delta n_s$ as a function of $\mchi$ for fixed $\Gamma_\chi = 10^{-12}$~GeV for two reheating temperatures. Here we see clearly that for 
$\trh = 10^5$~GeV the curvaton correction, $\delta n_s$ is small for all masses considered and therefore consistent with 2$\sigma$ of the observation. In contrast, for $\trh = 10^{13}$~GeV, the curvaton correction is large for $\mchi \gtrsim 3 \times 10^{8}$~GeV. Similarly, in Fig.~\ref{dnsgamma} we show a vertical cut on the $(\mchi,\Gamma_\chi)$ plane with fixed $\mchi = 10^9$~GeV as a function of $\Gamma_\chi$. In this case, we see that the correction to $n_s$ is excessive at small $\Gamma_\chi$.

For the new inflation model, the predicted value of $n_s$ is always below the observed value and a curvaton correction is needed to keep this model viable. In this case, therefore, the band corresponds to the $\pm 2 \sigma$ variation in $\Delta n_s$ shown in Fig.~\ref{nsprediction}. The allowed region for $\trh = 10^9$~GeV shown in the lower left panel of Fig.~\ref{fig2} is found at very low values of $\Gamma_\chi$ and masses of order $10^6 - 10^9$~GeV.  The ranges in $\Gamma_\chi$ and $m_\chi$ are larger for $\trh = 10^{13}$~GeV as seen in the lower right panel of Fig.~\ref{fig2}.

By comparing the gray bands in the upper and lower panels, we confirm our expectation that the larger $|\Delta n_{s}|$ contributed by the curvaton in the new inflation model requires larger $R_{\chi}$ and thus a later decay of $\chi$ (i.e., smaller $\Gamma_{\chi}$) for a given value of $m_{\chi}$.\footnote{Note that $H_{\rm end}$ of the Starobinsky model is 2 orders of magnitude larger than that of the new inflation model. This makes 
$P_{S_{\chi}}$ in (\ref{eq:PSchi}) of the two models comparable for the region of $m_{\chi}$ belonging to the gray bands. Thus, qualitatively $n_{s}-n_{*}$ in (\ref{eq:netns}) is mostly sensitive to $R_{\chi}$.}

We also show in Fig.~\ref{fig2} contours of constant $f_{NL}^{(\rm loc)}$ using purple (green) solid and dashed lines corresponding to $f_{NL}^{(\rm loc)}=9$ and $0$ ($f_{NL}^{(\rm loc)}=-1$ and $-11$) respectively. 
The contour with $f_{NL}^{(\rm loc)}=0$ runs close to (but slightly below) the MD/RD boundary with $x=1$. From Eqs.~(\ref{eq:parameter}) and (\ref{Rofx}) we find that  $f_{NL}^{(\rm loc)}=0$ corresponds to $x \simeq 1.85$.
Given the current constraint $f_{NL}^{(\rm loc)}=-0.9\pm5.1$~\cite{Planck:2019kim} at $68\%$ C.L., the regions in $(m_{\chi},\Gamma_{\chi})$ space surrounded by the purple solid and green dashed lines are excluded at the $2\sigma$ level. After applying this exclusion, we still see that significant fraction of each gray band remains viable, enabling a better fit to the observed $n_{s}$. Intriguingly, when $\chi$ decays in a RD era, it can produce $f_{NL}^{(\rm loc)}$ as large as $\mathcal{O}(1)$ in the regions near the intersection between the gray bands and the purple solid lines. If a large value of $f_{NL}^{(\rm loc)}=\mathcal{O}(1)$ is detected in a future survey, such regions will become of particular interest. On the other hand, the parts of the gray bands residing in the blue or cyan regions are characterized by $|f_{NL}^{(\rm loc)}|<1$ and $f_{NL}^{(\rm loc)}<0$. The associated curvaton decays in a MD era and characterized by smaller values of $m_{\chi}$ and $\Gamma_{\chi}$. Since $f_{NL}^{(\rm loc)}$ in (\ref{eq:parameter}) becomes negative for $x\geq  1.85$, the curvaton decaying in a MD era tends to produce a negative $f_{NL}^{(\rm loc)}$. Although the absolute magnitude$|f_{NL}^{(\rm loc)}|$ itself is small, the sign of $f_{NL}^{(\rm loc)}$ can be still invoked to distinguish a curvaton that decays in a MD era from one that decays in a RD era.

\begin{figure}[ht!]
  \centering
\includegraphics[width=0.45\textwidth]{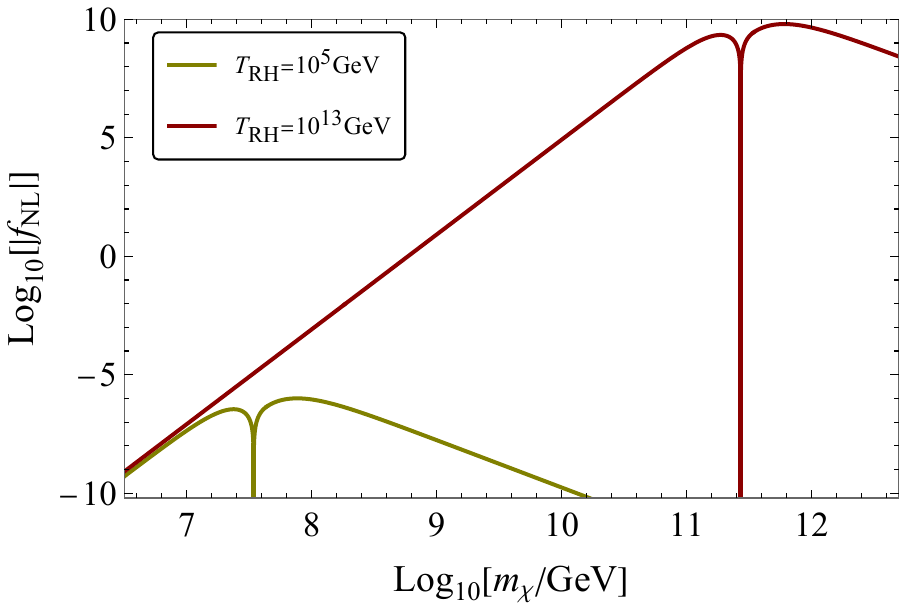}
  \caption{\em \small {$f_{\rm NL}$ as a function of $m_\chi$ for the Starobinsky model, with fixed $\Gamma_\chi=10^{-12}~{\rm  GeV}$ for $\trh = 10^5$~GeV (red) and $10^{13}$~GeV (green).
} 
}
  \label{fnlmchi}
\end{figure}

\begin{figure}[ht!]
  \centering
\includegraphics[width=0.45\textwidth]{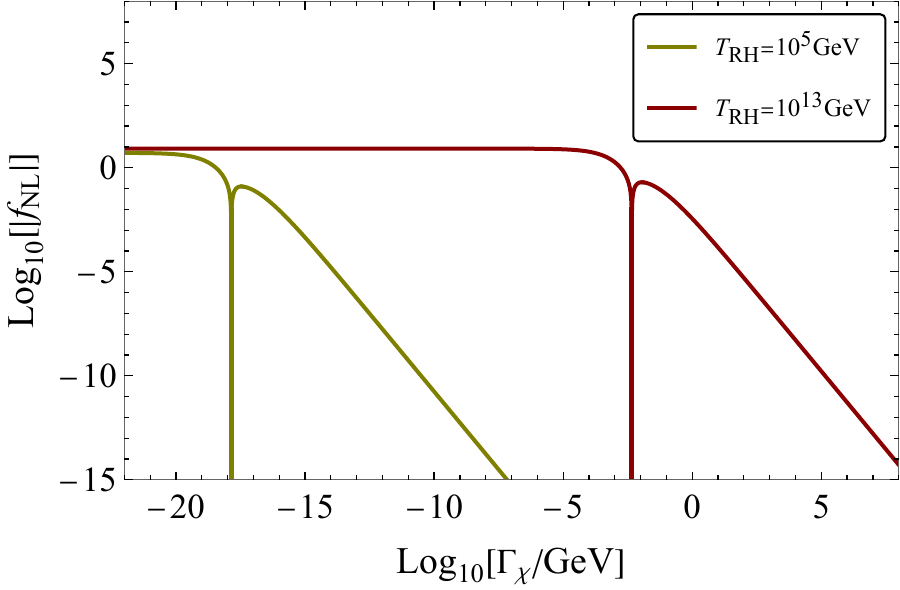}
  \caption{\em \small {$f_{\rm NL}$ as a function of $\Gamma_\chi$ for the Starobinsky model, with fixed  $m_\chi=10^{9}~{\rm  GeV}$ for $\trh = 10^5$~GeV (red) and $10^{13}$~GeV (green).
} 
}
  \label{fnlgamma}
\end{figure}

The behavior of $f_{NL}^{(\rm loc)}$ across the $(\mchi,\Gamma_\chi)$ plane is more clearly seen in Figs.~\ref{fnlmchi}
and \ref{fnlgamma}. In the former, we show 
$f_{NL}^{(\rm loc)}$ as a function of $\mchi$ for fixed $\Gamma_\chi = 10^{-12}$~GeV. For $\trh = 10^{5}$~GeV, the value of $f_{NL}^{(\rm loc)}$ is always small and remains within the Planck bound. However for $\trh = 10^{13}$~GeV, we see a rapid increase in $f_{NL}^{(\rm loc)}$ excluding masses $\mchi \gtrsim 10^9$~GeV.  Note the downward spike in $f_{NL}^{(\rm loc)}$ occurs when $x = 1.85$
when $f_{NL}^{(\rm loc)}$ changes sign (we are plotting the log of the absolute value). The analogous cut showing $f_{NL}^{(\rm loc)}$ as a function of $\Gamma_\chi$ for fixed $\mchi = 10^9$~GeV
is shown in Fig.~\ref{fnlgamma}.

Note that for the parts of gray bands residing in blue or cyan backgrounds, $\delta n_{s}$ and $f_{NL}^{(\rm loc)}$ are insensitive to a change in $\Gamma_{\chi}$ for fixed $m_{\chi}$. This is attributable to the fact that $R_{\chi}$ is almost constant when $x\gg1$. It is then difficult to infer $\Gamma_{\chi}$ given the insensitivity of $\delta n_{s}$ and $f_{NL}^{(\rm loc)}$ to $\Gamma_{\chi}$. 

We would also like to point out that
with the minimal cosmological assumptions,  the primordial (inflationary) gravitatioanl wave spectrum can be used for exploring curvatons residing in the intersections between gray bands and blue or cyan shaded regions. 
As a result of the entropy released from the late time decay of $\chi$ in a MD era, when compared to the modes re-entering the horizon after $\chi$-decay, the primordial gravitational wave spectrum of modes that reentered the horizon before $\chi$-decay is suppressed by $x=(s_{f}/s_{i})^{4/3}$ ~\cite{Nakayama:2008ip,Nakayama:2009ce,Kuroyanagi:2011fy,Jinno:2014qka,Kuroyanagi:2014nba,DEramo:2019tit,Choi:2021lcn}. In addition, the frequency $f_{\rm pGW}$ of the primordial gravitational wave spectrum at which the suppression occurs is given by
\be
f_{\rm pGW}\simeq2.65{\rm Hz}\left(\frac{g_{*}}{106.75}\right)^{1/2}\left(\frac{g_{*,s}}{106.75}\right)^{-1/3}\left(\frac{T_{d\chi}}{10^{8}{\rm GeV}}\right)\,,
\ee
where $T_{d\chi}$ is the temperature of the thermal bath when $\chi$ decays and obtained from $(3/2)H\simeq\Gamma_{\chi}$, and $g_{*}$ and $g_{*s}$ are the effective relativistic degrees of freedom for the energy density and entropy density. For $\Gamma_{\chi}\lesssim10^{-2}{\rm GeV}$ corresponding to $f_{\rm pGW}\leq\mathcal{O}(1){\rm Hz}$, such a suppression might be probed by the future spaced-based interferometers like LISA~\cite{LISA:2017pwj}, DECIGO~\cite{Seto:2001qf,Kawamura:2006up}, BBO~\cite{Crowder:2005nr,Corbin:2005ny,Harry:2006fi}, and future pulsar timing array like SKA~\cite{Carilli:2004nx,Janssen:2014dka,Weltman:2018zrl}.  Although this signature might be difficult to be seen for the new inflation model case due to the extremely small tensor-to-scalar ratio $r\simeq 2.4\times 10^{-8}$, it might be tested for the Starobinsky model featured by the relatively high $r\simeq 3.5\times 10^{-3}$.

Before concluding this section, we would like to note that the effect of the curvaton is not limited to either being constrained (as in the Starobinsky model) or increasing $n_s$ (as in new inflation). It is well known that the prediction of $r$ in chaotic inflation \cite{chaotic} is in excess of the current experimental limits on $r$ \cite{BICEP2021,Tristram:2021tvh}. However, from Eq.~(\ref{deltar}), we see that if the curvaton provides a significant contribution to the energy density before it decays, $\omega_\chi \lesssim 1$, and $\delta r$ may be large enough to bring the calculated value into agreement with experiment \cite{McDonald:2003xi,Sloth:2014sga,Fujita:2014iaa,Torrado:2017qtr,Sharma:2019qan,Suresh:2021bol}. In this case, $n_s$ is affected, and we must check whether or not consistency is maintained. For example, for quartic chaotic inflation,
For $N_* = 55$, $r \simeq 0.29$ and $n_s = .945$. Increasing $N_*$ to $\simeq 444$ will sufficiently lower $r$ to say, 0.036, but at the expense of increasing $n_s$ to 0.993. The inclusion of a curvaton can bring $r$ to 0.036
and requires only $N_* > 47$ to get $n_s > 0.956$. To avoid solutions with small $\Gamma_\chi$ in conflict with BBN, we find solutions with $\trh \gtrsim 10^5$~GeV  which can be achieved e.g., with $\mchi = 10^9 $ and $\Gamma_\chi \gtrsim 2.2 \times 10^{-21} $~GeV. In this case, $N_*=50.9$, $n_s=0.959$, $r=0.036$. 
Similarly for the quadratic model, 
for $N_* = 55$, the predicted values of $n_s$ and $r$ are 0.964 and 0.14, respectively. 
The curvaton would allow $r = 0.036$ and $n_s < 0.973$ but only if $N_* < 44$ in which case, $\trh < 7$~MeV 
and is close to being in violation of BBN constraints.

\section{Discussion}
\label{sec:disc}

The theory of inflation is an important element of what has become known as the standard cosmological model. Generic predictions include flatness ($\Omega = 1$), and the production of density perturbations which seed the formation of structure in the Universe.  Specific models of inflation also make predictions on the tilt of the anisotropy spectrum, $n_s$, the scalar-to-tensor ratio, $r$, and the degree of non-Gaussianity in the spectrum $f_{\rm NL}$. While $n_s < 1$, is a common prediction in models, its precise value may vary. Similarly, the range of predicted values of $r$ is greatly model dependent. In contrast, very small $f_{\rm NL}$ is a common feature of single field models. 

The precision at which $n_s$ has been measured \cite{Planck:2018jri,Planck} and the improvement in the limit on $r$ \cite{BICEP2021,Tristram:2021tvh} enables one to discriminate between different models of inflation. For example, models of new inflation (such as that studied here) predict a value of $n_s$ far outside 
the experimental range. Models of chaotic inflation predict a value of $r$ in excess of the observed upper limit, whereas the Starobinsky model \cite{Staro} predicts a 
value of $n_s$ is in very good agreement with experiment and a value of $r = 0.0035$ which is about a factor of 10 below the current upper limit. This may be tested in the next round of precision CMB experiments.
The improved limits in non-Gaussianity \cite{Planck:2019kim} have given further credence to single field inflation models. Thus a single field model such as the Starobinsky model, remains on solid footing as a candidate for this important part of our cosmological history. 

However, the physics of the early Universe may be more complicated than that described by a single field inflationary model.  The presence of a massive, unstable, scalar field may dramatically alter the predictions of a given model.
This scalar, the curvaton, may, upon decay,
transfer its isocurvature fluctuations (produced during inflation) to adiabatic fluctuations in the radiation produced by its decay. 

Here, we have followed the formalism in \cite{Fonseca:2012cj} to compute $\delta n_s$ and $\delta r$. For the two models of inflation considered here, $\delta n_s > 0$. For the Starobinsky model, we require the correction to be small (particularly when the reheating temperature is large). In contrast, as a single field inflation model, the new inflation model considered here is excluded, but with a correction $\delta n_s \sim 0.05$ may again be a viable candidate for inflation.

Within the context of these two models, we
have considered the the parameters $\mchi$ and $\Gamma_\chi$ which define the curvaton sector. And we have delineated the allowed parameter space which depends on the inflaton reheating temperature, $\trh$. 
For the Starobinsky model, since a correction is not absolutely necessary, there are a wide range of acceptable parameters, particularly those with large decay rates. In this case, the curvaton decays before it has a chance to dominate the energy density of the Universe and leave an imprint on the CMB spectrum. Nevertheless, we have also identified the range of parameters (seen in Fig.~\ref{fig2}) for which the curvaton corrections spoils the good agreement of the model. This is seen at relatively large curvaton masses and long lifetimes. 

In contrast to the Starobinsky model, the model of new inflation needs a curvaton to regain viability. Depending on the reheating temperature, there is only a narrow band of parameters where this is possible. For a somewhat low reheating temperature of $10^9$~GeV, the curvaton decay rate must be very small $\Gamma_\chi < 10^{-18.5}$~GeV  with a narrow range in curvaton masses. The parameter range increases at $\trh = 10^{13}$~GeV, but still $\Gamma_\chi \lesssim 10^{-12}$~GeV is necessary.

Finally, curvaton models are able to predict much larger amounts of non-Gaussianities than single field models. Though $f_{\rm NL} = 0$ is possible (parameters for this are found near the MD-RD boundaries for both cases II and III), most of the regions where $f_{\rm NL}$ exceeds the current Planck bounds occur where $\delta n_s$ is also excessive.  Nevertheless, there are some sets of parameters which provide an acceptable correction to $n_s$ and yet a value of $f_{\rm NL}$  which is large (and positive). 

Clearly the questions concerning the existence of a curvaton or a particular model of inflation requires new observations. More precision in $n_s$ 
may point to the existence of a curvaton even for the Starobinsky model. Of course non-zero determinations of either $r$ or $f_{\rm NL} $ would be greatly welcomed. 

\section*{Acknowledgements}
This work was supported in part by DOE grant DE-SC0011842 at the University of Minnesota.

\section*{Appendix A: Inflaton potentials from supergravity}
Both of the models of inflation we consider can be derived in the context of no-scale supergravity \cite{no-scale}
defined by the common K\"ahler potential
\beq
K = -3 \ln \left(T+T^{\dagger} - \frac13 |\varphi|^2 \right)
\eeq
where $T$ is the volume modulus chiral superfield and $\phi$ is the chiral superfield for the inflaton. The inflaton field is identified with the real part of the scalar component of $\phi$. Above, we have used $\phi$ for the scalar component of the chiral superfield for notational simplicity. The two models, new inflation~\cite{eenos} and Starobinsky~\cite{eno6}, are distinguished by the following different superpotentials 
\begin{equation}
W =  M \left(\varphi - \frac{\varphi^3}{9M_P^2}-\frac{\varphi^4}{\sqrt{2} M_P^3} \right) ~ {\rm new~inflation}  \label{Wnew}    
\end{equation}
\begin{equation}
W =  M \left( \frac{\varphi^2}{2} - \frac{\varphi^3}{3\sqrt{3} M_P} \right) \qquad {\rm Starobinsky} 
    \label{Wstaro}    
\end{equation}
Because of the non-trivial field-space manifold, a canonically normalized inflaton, $\varphi$, can be defined by 
\beq
\varphi = \sqrt{3} \tanh\left(\frac{\phi}{\sqrt{6}M_P}\right) \, .
\eeq
For the new inflation model, along the direction $\phi=\phi^{\dagger}$, the scalar potential in supergravity (SUGRA) leads to\footnote{This superpotential differs slightly from that in \cite{eenos}, as a different basis is being considered here. Here, $T$ is assumed fixed, whereas in \cite{eenos}, it was $K$ that was assumed fixed. The resulting potentials are very similar and Eq.~(\ref{eenospot}) was derived in the latter case. Eq.~(\ref{Wnew}) provides some higher order corrections to $V$ in (\ref{newpotential}) which do not affect the inflationary dynamics.}:
\begin{eqnarray}
    V_n(x)  = 
    & M^2 M_P^2 \left( 1-6\sqrt{6}\sinh \left(\frac{z}{\sqrt{6}}\right)^2\tanh \left(\frac{z}{\sqrt{6}}\right)\right)^2 
    \label{newpotential0}\\
     \simeq & M^2 M_P^2 (z^3-1)^2 \label{eenospot0}
\end{eqnarray}
where $z=\phi/M_P$ is the inflaton field value normalized by the Planck mass. \footnote{A similar model can be derived in the context of minimal supergravity (with $K = |\phi|^2$ and $W = (1-\phi)^2$ \cite{hrr}). Though it is difficult to obtain light scalar fields during inflation in minimal supergravity \cite{drt}, this problem can be avoided in no-scale supergravity \cite{GMO}. We do not attempt to construct a full curvaton model in supergravity here.}

The Starobinsky potential is given by:
\be
V_S(\phi)=\frac{3}{4} M^2 M_P^2  \left(1-e^{-\sqrt{\frac23}z}\right)^2\,.
\label{Spotential}
\ee
Though there are many other forms of the superpotential (\ref{Wstaro}) which lead to the same scalar potential \cite{eno7,enov}.

\section*{Appendix B: Corrections to $Ht$}
\label{sec:appB}
In section \ref{sec:scenarios}, we have defined the moment when $\chi$ oscillations start and when $\chi$ decays, by the equalities $\Bar{\beta} H=m_\chi$ and $\Bar{\beta}H=\Gamma_\chi$, where $\Bar{\beta}=\frac{3}{2}$ or $2$ for matter or radiation dominated scenarios. In reality, these relations receive a correction when the matter and radiation energy densities are comparable. 

To derive the exact relations between $H$ and $m_\chi,\Gamma_\chi$, let us define $Ht=1/\beta_x$ near $a_{\text{eq}}$ and $Ht=1/\beta_y$ near $a_{\text{RH}}$. In the former case, the total energy density is composed of radiation and $\chi$, with:
\begin{equation}
    \rho_{\text{tot}}=\frac{\rho_{\text{eq}}}{2}\left[\left(\frac{a_{\text{eq}}}{a}\right)^4+\left(\frac{a_{\text{eq}}}{a}\right)^3\right] \, ,
\end{equation}
which implies for the Hubble constant:
\begin{equation}
    H\equiv \frac{\dot{a}}{a}=\frac{\sqrt{\rho_{\text{eq}}}}{\sqrt{6}M_P}\left(\frac{a_{\text{eq}}}{a}\right)^2\sqrt{1+\frac{a}{a_{\text{eq}}}} \, .
\end{equation}
Integrating the above expression and using the relation $x=a_{d\chi}/a_{\text{eq}}$, we obtain:
\begin{equation}
t_{\text{eq}}=2M_P\sqrt{\frac{2}{3\rho_{\text{eq}}}}(2-2\sqrt{1+x}+x\sqrt{1+x})\equiv f[x(t)] \, .
\end{equation}
The time derivative of the function $f$ yields:
\begin{equation}
    1=\frac{\sqrt{6}M_P x}{\sqrt{\rho_{\text{eq}}(1+x)}}\frac{\dot{a}_{d\chi}}{a_{\text{eq}}} \, ,
\end{equation}
which eventually leads to the expression for $\beta_x$:\begin{equation}
    \beta_x=\frac{3x^2}{2(x^2-x-2+2\sqrt{1+x})} \, .
\end{equation}
One recovers $\beta_x=2$ ($\beta_x=3/2$) for $x\ll 1$  ($x\gg 1$). At the boundary of MD and RD era (i.e.~$a_{d\chi}=a_{\text{eq}}$ or $x=1$) we have $\frac{3}{4}(1+\sqrt{2})$.

On the other hand, before $a_{\text{RH}}$, the radiation redshifts as $a^{-3/2}$ whereas the matter energy density redshifts as $a^{-3}$. Introducing $y\equiv a_\chi/a_{\text{RH}}$, we obtain analogously:
\begin{equation}
    \beta_y=\frac{3y^{3/2}}{4(1+y^{3/2}-\sqrt{1+y^{3/2}})} \, ,
\end{equation}
which becomes $3/2$ for $y\ll1$, whereas at the boundary of case II and III (i.e.~$y=1$), we have  $\beta_y=\frac{3}{8}(2+\sqrt{2})$. After reheating, the inflaton energy density decreases exponentially, thus  the functional form of $\beta_y$ depends on the inflaton decay rate. Given that $\beta_y$ converges very rapidly to $2$, we approximate it as:
\begin{equation}
\beta_y=    \frac{1}{8}(3\sqrt{2}-10)e^{1-y}+2 \, .
\end{equation}


\end{document}